\documentstyle[11pt,leqno]{article}

 \newcommand{\beq}{\begin{equation}}                       
 \newcommand{\eeq}{\end{equation}}                         
 \newcounter{nt}[section]                                  
 \newcounter{nl}[section]                                  
 \date{}                                                   
 \textwidth  
12.5cm

\begin{document}

\vspace*{14mm}
\begin{center}
{ \Large \bf Non linear travelling waves with diffusion}
\end{center}

\vspace{2mm}
\begin {center}
{\normalsize Nota di M.De Angelis, E.Mazziotti}
   \footnote{ Facolt\`{a} di
             Ingegneria, Dip. Mat. Appl. "R. Caccioppoli",
 via Claudio 21, 80125, Napoli. E-mail: modeange@unina.it}
\end{center}

\begin {center}
{\small Presentata dal Socio  Pasquale Renno

(Adunanza del 3 Marzo 2006)}

\end{center}

\vspace{4mm}\noindent
{\small \it Key words:} {\small   Viscoelastic models, Superconductivity, Boundary-layer, Partial dif  \hspace*{20mm}ferential equations }

\vspace{8mm}\noindent
{ \small \bf Abstract-}{\small \@ A third order  parabolic operator $ \, {\cal L}_\varepsilon ,\,$   typical of  a non linear wave operator $ \, {\cal L}_0 \,$  perturbed by viscous terms, is analyzed. Some particular solutions related to $ \, {\cal L}_0 \,$ are explicitly determined and the initial value problem for $ \, {\cal L}_\varepsilon \,$ is considered. The parabolic-hyperbolic behaviour is analyzed and rigorous approximations of the solution are achieved.}

\vspace{8mm}\noindent
{ \small \bf Riassunto-}{\small  \@  Si prende in esame un operatore parabolico del terzo ordine $\ { \cal  L}_\varepsilon $ tipico di perturbazioni introdotte su equazioni iperboliche non lineari. Si ricavano anzitutto classi di soluzioni dell'equazione   non pertubata e si considera per}  $\ { \cal  L}_\varepsilon $ {\small il problema di valori iniziali in tutto lo spazio. Mediante una opportuna equazione integrale si determinano stime utili alla  valutazione rigorosa di approssimazioni di strato limite.}

\newpage
 \section{  Introduction}
 \setcounter{equation}{0}

\hspace{5.1mm}

\vspace{3mm}

   In a vast number of realistic mathematical models (ecology, superconductivity, spread of epidemics, neurobiology, viscoelasticity),
 the evolution is characterized by deep interactions between wave propagation and diffusion. These models are generally given by non linear hyperbolic equations perturbed by viscous terms described by higher - order derivatives with small  diffusion coefficients $ \, \varepsilon. \, $  Such small parameters imply very long times for the transmission of signals,  at least when the diffusion is the principle process involved.

In order to have an enough knowledge of the dynamic response, the order of diffusion effects and the time - intervals where the wave behavior prevails must be estimated. The valuation of these aspects leads to the analysis of non linear parabolic - hyperbolic boundary layers. A typical example is related to equations such as:

  \vspace{3mm}
\begin {equation}    \label {11} 
 {\cal L}_\varepsilon \,\,u \,\equiv \,[\,\varepsilon \partial_{xxt}
+ c^2\, \partial_{xx}  - \partial_{tt}  -  a \partial_t \,]\,u \, = \, f(x,t,u,u_x,u_t)
\end{equation}

\vspace{3mm}\noindent where $ \,u=u(x,t)\, $ and $\, \varepsilon,\, a,\, c\, $ are positive constants.

The parabolic equation (\ref{11}) includes models of the viscoelasticity \cite{r}-\cite{h} and  biology \cite{M}. In particular, when $ \, f\, = \, \, \gamma \, + sen\, u, \, \,$ one has the perturbed sine - Gordon equation (PSGE)which is basic in superconductivity \cite{bp},\cite{ls}.

In this paper, the initial value problem $\,{\cal P}_\varepsilon \,$ for the PSGE in all of the space is considered and the fundamental solution $ \, K_\varepsilon (x,t)\, $ of the linear operator $\,{\cal L}_\varepsilon $\, is explicitly determined. So, when $\, f\, $ is {\it linear } $\, ( f= f(x,t)),\,$ the problem $\,{\cal P}_\varepsilon \,$ is quite solved. More, $ \,K_\varepsilon \,$ is a $\, C^\infty \,$ function endowed with properties like those of the fundamental solution of the heat operator \cite{c}. For this, when $\, f\, $ is {\em not linear}, it is convenient to reduce the problem $\,{\cal P}_\varepsilon \,$to an integral equation whose kernel is just  $ \,K_\varepsilon \,$.

The qualitative analysis of this equation allows not only existence and uniqueness results, but rigorous estimates when $\, \varepsilon \, \rightarrow \, 0\, $ too. Let $\,{\cal L}_0 \,$ the operator to which $\,{\cal L}_\varepsilon \,$ is reduced for $\, \varepsilon \, \equiv \, 0\,$ and let $\,w\,$ the solution of the following initial value problem $\,{\cal P}_0: \,$ 

\vspace{3mm}
\begin {equation}    \label {12} 
 {\cal L}_0\,\,w \, = \, \bar f(x,t,u)\,\,\,\,\ with \,\,\ w(x,0)=f_0(x), \,\,\,w_t(x,0)=f_1(x). 
\end{equation}

\vspace{3mm}\noindent In the superconductive case, it results \cite{scott}
\vspace{3mm}
\begin {equation}    \label {13} 
 f(u)-\bar f(w)= sen \,w - sen\, u
\end{equation}

\vspace{3mm}\noindent and the {\it error} $\, v \, = u \,-\, w\,$ related to the approximation of $\, u\, $ by means of the wave solution $ \, w,\,$  is the solution of a problem $\,{\cal P}_\varepsilon \,$ with initial data vanishing and a source term given by

\vspace{3mm}
\begin {equation}    \label {14} 
 F_w(v) = sen\, w - sen \,u - \varepsilon w_{xxt}.
\end{equation}

\vspace{3mm}Then, according to several classes of special initial data $ \, (f_0, \, f_1,),\,$the solution $\,w\,$ of the reduced problem $\,{\cal P}_0\,$ is explicitly determined and, in corrispondence, the remainder term $ \, v\,= \,u\,-\,w\,$ is rigorously estimated, together with the time - interval where the diffusion effects are neglegible for $\, \varepsilon \, \rightarrow \, 0.\, $ As meamingful result one can see that whatever a positive constant $\, k\,<\,1\,$ is prefixed, the evolution is characterized by  the travelling wave $\, w\, $ and by diffusion effects  of the order of $ \, \varepsilon ^k,\,\,$ in each interval- time $ \, [\, 0, \, T_\varepsilon \,],\,\,$ with $\, T_\varepsilon \, $ defined in (\ref{612}).

\hspace{5.1mm}

\vspace{5mm}

\section{Statement of the problem}       

\setcounter{equation}{0}

\hspace{5.1mm}

\vspace{3mm}

If $T$ is a positive constant and

\vspace{4mm}

$\ \ \ \ \ \ \ \ \ \ \ \  \ \ \ \ \ \ \   \Omega_T =\{(x,t) :  x \in R
, \  \ 0 < t \leq T \}$,

\vspace{3mm}
\noindent
  let consider the following initial value problem ${\,\cal
P}_\varepsilon \,$:  

\vspace{3mm}
  \beq                                                     \label{21}
  \left \{
   \begin{array}{lll}
    & \partial_{xx}(\varepsilon
u _{t}+c^2 u) - \partial_t(u_{t}+au)=  f(x,t,u), &
        (x,t) \in \,\, \Omega_T,\\
\\
   & u(x,0)=f_0(x), \  \    u_t(x,0)=f_1(x), \ & x\in  R,  \\
\\
    
   \end{array}
  \right.
 \eeq

\vspace{3mm} \noindent where $\, f, \,f_0,\, f_1\,\,\,\, $   are arbitrary specified functions, and $ \,\,\, \varepsilon, \, a, \, c ,\,$ are positive constants.

When $\varepsilon \equiv 0$, the parabolic equation $(\ref{21})_1$ turns into the hyperbolic telegraph equation

 \vspace{3mm}
\begin{equation}                               \label{22} 
  (c^2 \, \partial_{xx}  - \partial_{tt} -\, a  \,\partial_t ) w = \bar f(x,t,w)
\eeq

\vspace{3mm}\noindent 
and the problem   ${\cal
P}_\varepsilon $ changes into a problem   ${\cal
P}_0 $ for $ w(x,t)$ which has {\em the same initial conditions  of }$\,\,{\cal
P}_\varepsilon. \,$\vspace{3mm}

\noindent When $\, \varepsilon \,$ is neglegible, a possible boundary - layer region could appear for $ \,T= \infty.$

\vspace{3mm}

In order to estimate the influence of the dissipative term $\, \varepsilon \, u_{xxt}\,$ on the wave behavior of $\,w\,$, the difference 

\vspace{3mm}

\begin{equation}                \label {23}
 u(x,t,\varepsilon)  - w(x,t)  = v(x,t,\varepsilon)
\end{equation}

\vspace{3mm}\noindent
is to be evaluated.  For this, one has the  following  problem  ${\,\cal
P}_v\,$   related to the {\em  remainder} term $ \, v :\,$

\vspace{3mm}
  \beq                                                     \label{24}
  \left \{
   \begin{array}{lll}
    \varepsilon \, v_{xxt}+c^2\, v_{xx} -v_{tt}-av_t  =  F_w(x,t, v )   & (x,t) \in \,\, \Omega_T, \\
\\
 v(x,0)=0, \  \   v_t(x,0)=0, \  & x \in R,\vspace{2mm}  \\
\\
   
   \end{array}
  \right.
 \eeq 

\vspace{3mm}\noindent where the source term $ \, F_w \, $ is given by:

\vspace{3mm}
\beq                   \label{25}
\,\, F_w(x,t,v)= f(x,t,w+v) -\bar f(x,t,w) - \varepsilon \,{w}_{xxt}.\,\,
\eeq

\hspace{5.1mm}

\vspace{5mm}

\section{Linear case : fundamental solution  and properties }       

\setcounter{equation}{0}

\hspace{5.1mm}

\vspace{3mm}

When $\,F_w = F(x,t)\, $ is linear and the Laplace transform is applied to the problem  (\ref{24}), the transform $\hat v(x,s)$ of the solution $v(x,t)$ is:

\vspace{3mm}

\begin{equation}                  \label {31}
\hat v(x,s) = \int_ R  \, 
 \hat{ K}_\varepsilon
(x - \xi,s) \ \hat F(\xi,s) \, d\xi,
\end{equation}

\vspace{3mm}\noindent where $\, \hat F \,$ is the transform of $\, F \, $ and

\vspace{3mm}

\begin{equation}                           \label{32}
\hat K_\varepsilon (x,s) = \
\frac{e ^{ \,-\, |x| \, \sqrt {s(s+a ) / (\varepsilon s + c^2) }}}{ 2 \, \sqrt{ s (s+a) (\varepsilon
s +c^2)}}.
\end{equation}

\vspace{3mm}

To obtain the inverse Laplace tranform $\,{\cal L  }_t^{-1}\,$ of the function (\ref{32}), let  $r = |x|/\varepsilon, \,\, b= c^2/\varepsilon $ and

\vspace{3mm}

\begin{equation}                           \label{33}
\hat G \,(r,s) = \
\frac{e ^{ \,-\, r \, \sqrt {s(s+a ) / ( s + b) }}}{ 2 \,  \sqrt \varepsilon  \, \sqrt{  (s+a) (s +b)}},
\end{equation}

\vspace{3mm}
\noindent 
so that:

\vspace{3mm}

\begin{equation}                           \label{34}
\hat K_\varepsilon  \,(x,s) = \
\frac{1}{\sqrt s}\, \, \hat G ( r,s) \, = {\cal L  }_t\,\,\, [\,\frac{1}{\sqrt{\pi\,t}}\, \, \ast G(r,t)\,].
\end{equation}

\vspace{3mm}
\noindent 
If   $ I_n (z)$ denotes the modified Bessel function, let

\vspace{1mm}   
\beq                 \label{35}
G(r,t) = \, \frac{r}{4 \, \sqrt{\pi \, \varepsilon}} \,\, \int^t_0 \,\, \frac{e^{- \frac{r^2}{4v}}}{  v\,\sqrt{v}}  \, \,\,e^{-\,b (t-v)}\,\, I_0 (r \sqrt{(b-a)(t-v)/v}\,\,\,dv.
\eeq

\vspace{3mm}

Then, the following theorem holds:

\vspace{5mm}

{\bf Theorem 3.1}- {\em For all  }$\,r>0,\,$  {\em the Laplace integral}  $\,{\cal L  }_t\,\,G(r,t)\,$ {\em converges absolutely
in the half-plane} $ \Re e  \,s > \,max(\,-a,\,-b\,),\,$ {\em and one has:}

 \vspace{3mm}

\beq      \label{36}
\,{\cal L  }_t \, G(r,t)\,\,= \,\hat G(r,s)\,\,,
\eeq

\vspace{3mm}\noindent
{\em with }$ \,\hat G
\, $ {\em and} $\, G \,$ {\em defined by }(\ref{33})-(\ref{35}).

\vspace{3mm} 
{\bf Proof -} By Fubini- Tonelli theorem it results

\vspace{3mm}

\beq                 \label{37}
{\cal L  }_t\,G  =  \frac{r}{4  \sqrt{\pi  \varepsilon}}  \int ^\infty_0   e^{- (sv+\frac{ r^2}{4v})} \frac{dv}{  v\sqrt{v}}  \int ^\infty_0 e^{-(s+b)\,z} I_0 (r \sqrt{(b-a)z/v  }) dz
\eeq

\vspace{3mm}\noindent and further \cite{e} it is:

\vspace{3mm}

\beq                 \label{38}
  \int ^\infty_0   e^{- (s+b) z} \,\, I_0\, (\,r \sqrt{(b-a)z/v}\,\,)\, dz \,\,= \frac{e^{\,\,\frac{r^2}{4v}\,\,\frac{b-a}{s+b}}}{s+b}.
\eeq

\vspace{3mm}\noindent By means  \cite{e} of known formulae, (\ref{37}) and (\ref{38}) imply (\ref{36}). The absolute convergence holds whatever the constants $\, a,\, b\, $ may be. In fact, when $ \, a>b,\, $ for all real $\, z \geq\, 0, \,$ one has:

\vspace{3mm}

\beq                 \label{39}
  |\, J_0 ( r \sqrt{(a-b)\,z}\,| \, \leq 1 \leq  I_0\, (\,r \sqrt{(b-a)\,z}\,\,).\,
\eeq
\hbox{}\hfill\rule{1.85mm}{2.82mm}

\vspace{5mm}   {\bf Remark 3.1 -} The function G(r,t) given by (\ref{35}) is defined also when  $ \,r\,=\, |x|\,/\,\sqrt {\varepsilon} \,=\,0.\,$ In fact,  if in the integral of (\ref{35})  one puts : $ \,r^2 \, (t-v)/v \,=\,z,\,$ one has

\vspace{3mm}

\beq                 \label{310}
G(r,t) = \, \frac{e^{- \frac{r^2}{4t}}}{4 \, \sqrt{\pi \, \varepsilon \,t}} \,\, \int^\infty_0 \,\, e^{- \frac{z}{4t} \,-\frac{z}{z+r^2}\,b\,t}  \,\,\,\,  \frac{I_0 (\, \sqrt{(b-a)\,z}\,\,)}{\sqrt{z+r^2}}\,\,\, dz
\eeq
\vspace{3mm} and for $r=0$, it results:

\vspace{3mm}
\beq                 \label{311}
G(0,t) = \, \frac{e^{- bt}}{4 \, \sqrt{\pi \, \varepsilon \,t}} \,\, \int^\infty_0 \,\, e^{- \frac{z}{4t}} \,\,I_0 ( \sqrt{(b-a)\,z} \,)\,\,\,\frac{dz}{\sqrt{z}}\,= 
\eeq
\hspace{4cm}\[  =\,\,\frac{1}{2\,\varepsilon} \,\, e^{-\,\frac{a+b}{2}\,t\,}\,\,\,I_0 (\,\frac{b-a}{2}\,t).\]

\vspace{3mm}\noindent
As consequence, by (\ref{34}) it follows

\vspace{3mm}

\begin{equation}                           \label{312}
K _\varepsilon  \,(x,t) \, = \, \int^t _0 \,  G( r,\tau) \, \,\frac{d\tau}{\sqrt{\pi (t-\tau)}}
\end {equation}

\vspace{3mm}\noindent
and, for $ \, x=0\,$, it is:

\vspace{3mm}
\beq                 \label{313}
K_\varepsilon (0,t) = \, \frac{1}{2}\,\, \int_0^t \frac {e^{\,-\,\frac{a+b}{2}\,\tau\,}}{\sqrt{\pi \, \varepsilon \,(t-\tau)}}\,\, I_0 (\,\frac{b-a}{2}\,\tau)\,\, d\tau. 
\eeq
\hbox{}\hfill\rule{1.85mm}{2.82mm}

\vspace{5mm}\noindent 
Moreover, it's easy to prove the following theorem:

\vspace{5mm}

{\bf Theorem 3.2}- {\em The function  }$K_\varepsilon $  {\em
defined by }(\ref{310})- (\ref{312}) {\em is   a  } $ \, C^\infty (\Omega_T) \,$ {\em solution of the equation } $\, L_\varepsilon \,u\,=\,0.\,$ {\em More, when }$ a<b=c^2/\varepsilon, \,\,\, K_\varepsilon \,$ {\em is  never negative  in } $ \Omega_T. \,$    
\hbox{}\hfill\rule{1.85mm}{2.82mm}

\vspace{5mm}\noindent
As (\ref{310})- (\ref{312}) show, the fundamental solution  $K_\varepsilon (x,{\bf \cdot}\,),\, $ qua function of x, is an even function which results positive when $\, a<b\, $ and so

\vspace{3mm}
\begin{equation}                  \label {314}
  0\,\leq\, \int_0^\infty  \,\,K_\varepsilon (x-\xi,t)\, \,d\xi \,\,\, = \, 2\,\int_0^\infty \,\,K_\varepsilon \,\,(y,t) \,dy.
\eeq

\vspace{3mm}\noindent More by  (\ref{32}) one has
 
\vspace{3mm}
\begin{equation}                  \label {315}
  0\,\leq\, \, 2\,\int_0^\infty \,\,\hat K_\varepsilon \,\,(y,s) \,dy \, = \, \frac{1}{s^2 +as}\, =\, {\cal L}_t \,[\,\frac{1}{a}\,(1-e^{-\,at})\,]
\eeq

\vspace{3mm}\noindent and the absolute convergence of the integral (\ref{36}) implies the following estimate:

\vspace{5mm}

{\bf Theorem 3.3}- {\em  When} $\, a\,\leq \, b\,$ {\em it results}

\vspace{3mm}
\begin{equation}                  \label {316} 
  0\,\leq\, \, \,\int_0^ \infty \,\, K_\varepsilon \,\,(x-\xi,t) \,d\xi \, = \,(\,\frac{1}{a}\ )(1-e^{-\,at})\,\leq \,1/a.
\eeq
\hbox{}\hfill\rule{1.85mm}{2.82mm}

\vspace{5mm}This property of $ \, K_\varepsilon \,$ is useful to various estimates for the non linear case.

\vspace{3mm}
\vspace{5.1mm}

\section{ Reduced equation and travelling waves}       

\setcounter{equation}{0}

\hspace{5.1mm}

\vspace{3mm}

As for the model of Superconductivity, the approximation  $\,w\,$ is characterized  by the source $\, \bar f \,= \, sen\, w \, +\, \gamma,\,\,\,$so  that one has

\vspace{5mm}
\beq                \label{41}
  w_{xx} \,- \,w _{tt} \, - \, \ a w_ {t} =   \sin w \,  \,+ \gamma  
\eeq

\vspace{3mm}\noindent 
where it's assumed $\,\, c^2 =1\,\,$ without loss of generality.

Let $\, \psi \,$ an arbitrary function and let $ \,\,\Pi (\psi)\,\,$  the Lobachevsky's angle of parallelism \cite{gr}, defined by

\vspace{3mm}

\beq     \label{42}
\left \{
   \begin{array}{lll}
\Pi(\psi) = \, 2 \, arctg\,\, e^\psi \,\, =\,\, 2 arctg \,\, e^{-\psi} & [ \psi\geq 0] \\
\\
\Pi(\psi) = \, \pi - \Pi (- \psi)  & [ \psi<0],
   \end{array}
  \right.
\eeq

\vspace{3mm}\noindent
and such that 

\vspace{3mm}
\beq                \label{43}
  sen \,  \Pi (\psi) \,\, = \frac{1}{cosh (\psi) }.   
\eeq

\vspace{3mm}
A class of travelling wave solutions of (\ref{41}) can be explicitly evaluated whatever $\, \gamma \, $ may be; it suffices to consider functions  $\,\,\psi \,=\, \psi (\xi)$  such as:

\vspace{3mm}
\beq                \label{44}
w \,=\,\Pi \,[\,\psi(\xi)\, ] \, = \, \varphi ( x , t), \,\,\,\,\,\,\ with \,\,\,\,\,\xi = \frac{1}{a} \,\, (x-t).   
\eeq

 \vspace{3mm}
Then  (\ref{41}) becomes 

\vspace{3mm}
\beq                \label{45}
  - \, \ a \varphi_ t =   \sin \varphi \,  \,+ \gamma  
\eeq

\vspace{3mm}\noindent and the function $ \, \psi \, $ is given by

\vspace{3mm}
\beq                \label{46}
\frac{d\psi}{1+\,\gamma \,cosh (\psi) } \,\,\ = \,\,d\xi.   
\eeq

\vspace{3mm} According  to all the  values of  $ \,\, \gamma \,\, $, indicating by k an  arbitrary constant of integration,  
 one has the following special solutions of the non linear equation (\ref{41}).

\vspace{3mm}\noindent
- {\em Case } $\, \gamma \, = \, 0\,$ 
\vspace{3mm}
\beq                       \label {47}
w(x,t) \,\ =  \, \Pi ( \xi \,+ \, k),  
\eeq

\vspace{3mm}\noindent
- {\em Case } $\, \gamma \, = \, 1.\,\, $ If $\,\,  \xi -  k\, =\bar \xi, $ one has:

\vspace{3mm}
\beq                       \label {48}
w(x,t) \,\ =  \, 2 \,\, arctg (\,- \,\frac{ \bar \xi  +2 }{ \bar \xi} ),
\eeq

\vspace{3mm}\noindent
- {\em Case } $\, \gamma ^2 \, < \, 1.\, \,$  If $ \,\alpha=\sqrt{1-\gamma^2}, \,\,\,$ $\,\,\,(\,\alpha \,/ \,k \,)\,\xi \,=\,\bar \xi, \,  \,\,$  one has: 

\vspace{3mm}
\beq                       \label {49}
w(x,t) \,\ =   \, 2\,\, arctg \, [ \,\, \frac{\alpha}{\gamma} \,\, ( \frac{1+ k\,\,e^{\,\,\bar \xi}}{1-  k\,\,e^{\,\,\bar \xi}} \, - \,\, \frac{1}{\alpha}\,\,)  \,\, ]. 
\eeq

\vspace{3mm}\noindent
- {\em Case }  $\, \gamma ^2 \, > \, 1.\, $  If  $ \,\, \sigma =\sqrt{\gamma^2-1},\,\,\,\,$$\,(\,\sigma \,/ \,2\,)\, \xi \, = \bar \xi, \, \,\,\,$  one has

\vspace{3mm}
\beq                       \label {410}
w(x,t) \,\ =    2 \,\,arctg \, \{ \,\, \frac{1}{\gamma} \,\,[ 1  +  \sigma\, tg \, ( \bar \xi\,+\, k)] \}. 
\eeq

\vspace{3mm}
\vspace{5.1mm}

\section{ An integral equation}       

\setcounter{equation}{0}

\hspace{5.1mm}

\vspace{3mm}

Consider now the non linear problem defined in (\ref{24}) and assume that the prefixed approximation $ w(x,t)$ is compatible with the following assumptions on the function $\, F_w \, $ given by (\ref{25}):

\vspace{3mm}\noindent

{\bf Conditions A }

\vspace{3mm}

1)  \hspace{2mm} $\,F_w(x,t,v)\, $ is defined and continuous on the set 

\vspace{3mm}

\begin{center}
 $ \Upsilon \,\,=  \{ \, (x,t,v)\, \in  \Omega_T \,\, \, \times \, (\, -\infty, \, + \infty)\,\}$
 
\end{center}

\vspace{5mm}

2)  \hspace{2mm} For each $\,\, k\,>0 \,\,$ and for $\, \,|v|\,<k\,\,$, the function $\, F_w(x,t,v) \,\,$ is uni-

\hspace{8mm}formly Holder continuous in $ \, x\,$ and $\, t\,$ for each compact subset of 

\hspace{8mm}$\, \Omega_T \,$  .

\vspace{5mm}

3)  \hspace{2mm} There  exists a constant $ \, C_F \, $ such that 

\begin {center}
$ \, | \,F_w\,(x,t,v_1) \, - \, F_w\,(x,t,v_2)\,| \, \leq \, C_F \,\, | \,v_1 -v_2\,|$
\end {center}

\hspace{9mm}holds for all $\, v_1, v_2\,$ and $\, w. \, $

\vspace{5mm}

4)  \hspace{2mm}  $\, F_w\, $ is bounded in $ \, \Omega _T \,$ for bounded $ v$ and $w.$

\vspace{5mm}\noindent

{\bf Definition 5.1  -}For each prefixed $w(x,t)$, by a solution of the problem $ {\cal P}_v$ we mean a function  $\,\,v\,\,\,\,$continuous and bounded in $ \Omega _T $
such that, whatever $ \varepsilon >0 $ may be, the derivatives $ \,v_t,\, v_{tt}, \,v_{xx}, \,v_{xxt}\,$  are continuous in  $ \Omega _T $ and verify (\ref{24}).

\vspace{5mm}
According to the formal representation (\ref{31}) it results:

\vspace{3mm}

\begin{equation}                  \label {51}
v(x,t) = \int ^t_0 \, d\tau \,\int_ R  \, 
 { K}_\varepsilon(|x - \xi|,t-\tau) \  F_w\,[ \,\xi, \,\tau,\, v(\xi, \tau)\,] \, d\xi
\end{equation}

\vspace{3mm}\noindent
 and the conditions A imply that, for a solution $ \,v(x,t)\,$ of the problem  $ {\cal P}_v, \,$the function $\,\, \lambda (\xi,\tau)\, = \, F_w\,[\,\xi, \,\tau,\, v(\xi, \tau)\,] \,\, $is bounded in   $ \,\Omega _T \,$ and is uniformly Holder continuous in $\,\xi \,$ and continuous in $\,\tau \,$ on each compact of $ \,\Omega _T .\,$

If

\vspace{3mm}

\begin{equation}                  \label {52}
|| \, F_w \, ||_t \,\,  = \,\,{\sup_ {0\leq \tau \leq t }}\,\,\ \sup _{x\in R}\,\, |\, \lambda (\xi,\tau)\,|,
\end{equation}

\vspace{3mm}\noindent by (\ref{51}) and theorem 3.3 one deduces

\vspace{3mm}

\begin{equation}                  \label {53}
|\,v(x,t)\, | \,\,  \leq \,\,
(t/a) \,\,\, || \,F_w \, ||_t \,  \,\,\, \\ \ \ \ \,\,\,\,\,\,\,\ \\ x\in R , \,\,\ t\geq 0 
\end{equation}

\vspace{3mm}\noindent and so, for each compact subset of$ \,\,\, -\infty  < x\,< \infty,\, \,\,\,\, 0\,<\,t \, \leq T,\, $ the function $\, v\, $ is uniformly Holder continuous. Consequently,  by Conditions A, the function $\, F_w \,[\,x,t,v(x,t)\,]\,\,$ is uniformly Holder continuous too. More, by means of the properties of the kernel $\, K_\varepsilon, \,$ it's possible the following statement.

\vspace{5mm}
{\bf Theorem 5.1 - } {\em For each prefixed }$ w$, {\em the initial-value problem} (\ref{24}) {\em admits a unique solution if and only if the integral equation  } (\ref{51}) {\em possesses a unique solution } $\, v\,\, ${\em  continuous and bounded for } $ \, (x,t) \, \in \, ( -\infty, + \infty) \, \times \, ({0,T)}.$
\hbox{}\hfill\rule{1.85mm}{2.82mm}

\vspace{5mm} Consider now a positive $\, \eta \, < \, T\, $ and let

\vspace{3mm}
\beq                \label {54}
|| \, v \, ||_\eta \,\,  = \,\,{\sup_ {0\leq t \leq \eta }}\,\,\ \sup _{x\in R}\,\, |\, \ v (x,t)\,| 
\eeq

\vspace{4mm}
\beq                 \label{55}
\nonumber
{\cal B}_\eta \, \equiv \, \{ \,v(x,t): v\in C(\,\,(-\infty, \infty )\times [\,0,\eta\,]\,\,)\,\, and \,\,||\,v \,\,\,||_\eta\,\ < \infty \,  \}
\nonumber
\eeq

\vspace{4mm}\noindent The set  $\,{\cal B}_\eta \,$ is a Banach space and the mapping

\vspace{3mm}
\beq                \label {55}
 {\cal F} \,v (x,t)\, \,\,  =  \int ^t_0 \, d\tau \,\int_ R  \, 
 { K}_\varepsilon(x - \xi,t-\tau) \  F_w\,[ \,\xi, \,\tau,\, v(\xi, \tau)\,] \, d\xi,\,\,
\eeq

\vspace{3mm}\noindent  owing to theorems 3.2 - 3.3, maps $\,{\cal B}_\eta \,$ into $\,{\cal B}_\eta. \,$ Then, by means of the well-known techniques of the fixed point, the following theorem can be proved.

\vspace{5mm}
{\bf Theorem 5.2 - } {\em When} $\, F_w (x,t,v)\,$ {\em satisfies the conditions A, then the initial- value problem } $\,  {\cal P}_v \, $ {\em given by } (\ref{24}) {\em admits a unique solution.} 
\hbox{}\hfill\rule{1.85mm}{2.82mm}

\vspace{5mm}
In conclusion, according to the prefixed initial data $ \, f_0 (x),\, f_1(x)\, $ of the problem ${\, \, \cal P}_\varepsilon, \, \,$ one can solve explictly the reduced problem ${\, \,\cal P}_0\,\,$ by means of the research of travelling waves outlined in sect.4 and by numerous class of solutions of sine- Gordon equations available in literature.

When the solution $\, w \, $ of ${\,\,\cal P}_0\,\,$ is determined, the source term $\, F_w \,$ is known and its properties can be analyzed in order to show that theorems 5.1 and 5.2 can be applied. Then, in this case, the integral equation (\ref{51}) allows to estimate the remainder term $\,v.\,$

\vspace{5.1mm}

\section{Estimate of the diffusion - An example }       

\setcounter{equation}{0}

\hspace{5.1mm}

\vspace{3mm}

For example, consider the approximation $\,w \,\,$ induced by the initial conditions

\vspace{3mm}
  \beq                                                     \label{61}
  \left \{
   \begin{array}{lll}

 w(x,0)= 2 arctg ( e^{x/a})  \\ 

\\
   w_t(x,0)= - \frac{2}{a} \,\, \frac{e^{x/a}}{1+e^{2x/a}}  
\\
   
   \end{array}
  \right.
 \eeq 

\vspace{3mm}\noindent
and let $\, \xi \, = \frac{1}{a} \, (x-t).\,$ Then, the problem  ${\, \,\cal P}_0\,\,$ given by (\ref{41}) - (\ref{61})admits the solution

\vspace{3mm}
\beq                       \label {62}
w(x,t) \,\ =  \, 2 \,\, arctg  \,( \,e^ {\xi }\, )
\eeq

\vspace{3mm}\noindent
which represents a travelling wave with speed $\, c\,=1\,$. It results:

\vspace{3mm}
\beq                     \label{63}
 w_{xxt}\, =\, - \, \frac{2}{a^3}\, \ e^\xi \,\, \ \frac{e^{4\xi}\,- 6\, e^{2\xi}\,+1}{(\, e^{2\xi}\, +1 \,) ^3} \,\\  \,\,\,\,\,\,\,\,\,\,\ \,\,\,(\xi \, \in R \, )
 \eeq

\vspace{3mm}\noindent
and so

\vspace{3mm}
\beq                     \label{64}
 \,|\,w_{xxt}\,| \,\leq \, \, \beta \,\,\ \ \ \ \ \ \ \ \ \ \forall \,(x,t) \,\in  \,\Omega_T
 \eeq

\vspace{3mm}\noindent where the constant $\,\beta \,$ depends only on $\, a\,$. As consequence, the source $ \, F_w\,(x,t,v)\, $ related to the problem  
${\, \,\cal P}_v\,\,$for the superconductive model is

\vspace{3mm}

\beq     \label{65}
 F_w(x,t,v) \, = sen \,\, (v+w) \,\, -sen \, w \,\, - \varepsilon  w_{xxt} 
\eeq

\vspace{3mm}\noindent and satisfies the conditions A. More, by (\ref{64})one has:

\vspace{3mm}

\beq     \label{66}
| \, F_w(x,t,v) \,|  \, \leq  \, |\, v(x,t,\varepsilon ) \, | \, + \, \varepsilon \, \beta. 
\eeq

 \vspace{3mm}

Then, by theorems 5.1 and 5.2, one deduces that the error $\, v\,$ related to the approximation $ \, u_\varepsilon \, \sim \, w\, \,$ is such that:

\vspace{3mm} 
\beq            \label {67}
|v(x,t,\varepsilon)|\, \leq \, \,\int ^t_0  d\tau \, \int_R  [\, \, |\,v(\xi,\tau,\varepsilon)\,|\,+\,\varepsilon \,\, \beta\,\, ] \, |\,K_\varepsilon (x-\xi, \,t- \tau)\,|\,d\xi. 
\eeq

  \vspace{3mm} \noindent When $\, \varepsilon \, \rightarrow \, 0, \, $ it  results $ \, a \,<\,b\, =\, \frac{c^2}{\varepsilon} \, $ and   $\, K_\varepsilon \, \geq \, 0\,$ according to theorem 3.2. If one puts

\vspace{3mm}
\beq  \label{68}
r_\varepsilon(t) \,\, = \,\,\sup_{R}\,\, |v(x,t,\varepsilon)|,
\eeq

\vspace{3mm}\noindent
by (\ref{67}) and theorem 3.3 it follows

\vspace{3mm} 
\beq    \label{69}
 0\, \leq \, r_\varepsilon (t)\,\, \leq \,\,\frac{1}{a} \\\, \int ^t_0 \,  r_\varepsilon (\tau) \,d\tau \,\, + \, \frac{\beta}{a} \, \,\varepsilon \, t\, 
\eeq

\vspace{3mm}\noindent
and so the Gronwall Lemma implies

\vspace{3mm} 
\beq     \label{610}
0\, \leq \, r_\varepsilon  (t)\,\,\, \leq \,\,\,[\, \, \beta \, (\,{T }/{a}\,) \, \, \, e^{\,T/a} \,\,] \,\, \varepsilon \ \ \ \,\,\,\,\ \ \ \ \ \,\,\, \forall \, t \, \in \,[0, T]. 
 \eeq

\vspace{3mm} This estimate allows to specify the infinite time - intervals where the effects of diffusion are of the order $\, \varepsilon ^k \, \,$ with $\, k\, <\,1 .$ In fact, whatever $ 0\,<\,k\,<\,1\,$ may be, let

\vspace{3mm}
\beq                 \label{611}
 \, T_ \varepsilon \, = \,\,\,\, \frac{a}{2}\, \,\,ln \, \,[\,\, \frac{1}{\beta} \,\,  \frac{1}{\varepsilon ^{1-k}}\,\,].
\eeq

\vspace{3mm}\noindent Then, for $\,T\,\leq \,T_\varepsilon,\,$ by (\ref{610}) it results

\vspace{3mm}  
\beq       \label{612}
0\, \leq \, r_\varepsilon (t)\, \leq\, \beta \, \,e^{\,2T/a\,} \,\, \varepsilon \,\, \leq \,\,\varepsilon ^k \,\,\,\,\,\forall t \, \in [0,T_\varepsilon].    
\eeq

\vspace{3mm}\noindent   

\vspace{3mm}So, {\em when} $\varepsilon \,$ {\em is vanishing, the evolution of the superconductive model  is characterized by  the travelling wave }$\, w\, $ {\em given in} (\ref{62}) {\em and by diffusion effects which are of the order of }$ \, \varepsilon ^k\,\,$ {\em in each interval- time }$ \, [\, 0, \, T_\varepsilon \,],\,\,$ {\em with} $\, T_\varepsilon \, $ {\em given by }(\ref{611}) {\em and} $\, k\,<\,1.\,$

\vspace{5mm}

\hspace{5.1mm}

\end{document}